\documentclass[twocolumn,floats,showpacs,prd,amssymb,floatfix,nofootinbib,balancelastpage,superscriptaddress,amsmath]{revtex4}
\usepackage{epsfig}
                              \newlength{\strikewidth}
                              \newlength{\strikelength}
\begin{document}

\title{Hunting the lightest lightest neutralinos}

\author{Stefano Profumo}
\email{profumo@scipp.ucsc.edu}
\affiliation{%
Santa Cruz Institute for Particle Physics and Department of Physics,\\ University of California, Santa Cruz CA 95064
}%

\begin{abstract}
\noindent  The lightest neutralino in the minimal supersymmetric extension of the Standard Model can be, in principle, massless. If super-light neutralinos are the dark matter, structure formation constrains their mass to be above a few keV. We show that relaxing the assumption of radiation domination and entropy conservation prior to Big Bang Nucleosynthesis, the relic abundance of very light neutralinos can be consistent with the inferred cold dark matter density. We study how we can hunt for light neutralino dark matter, with a mass at or below a GeV, focusing on both direct and indirect searches. We argue that the two most promising channels are spin-dependent direct detection and the search for monochromatic gamma rays from the prompt pair-annihilation of neutralinos into photons with GLAST. Our study indicates that the lightest lightest neutralinos can be detected as long as their mass is above a few tenth of a GeV, a mass range where a future linear collider could provide important information on the details of the particle dark matter model.

\end{abstract}




\maketitle

\section{Introduction}

The nature of particle physics beyond the Standard Model of strong and electro-weak interactions might be, potentially, deeply interwined with the issue of understanding at the fundamental level what non-baryonic dark matter is. Several frameworks exist where New Physics at the electro-weak scale is connected to the nature of particle dark matter. An outstanding example among these frameworks is supersymmetry: If supersymmetric particles exist at the electro-weak scale, a natural expectation is that the lightest supersymmetric particle (LSP) is related to the dark matter which comprises most of the matter content of the universe. Excitingly enough, the Large Hadron Collider (LHC) will soon shed light on what lies beyond the energy scales where the Standard Model has been tested so far. The LHC could thus initiate an era where the {\em Quantum} is directly connected to the {\em Cosmos} \cite{Kolb:2000tc} in the quest for the main ingredient of our own universe.

Within supersymmetry, neutralinos stand out as leading candidate dark matter particles. Neutralinos are the mass eigenstates resulting from the superposition of the fermionic super-partners of neutral gauge bosons and of the two neutral Higgses of the minimal supersymmetric extension of the Standard Model (MSSM). Neutralino LSP's have the remarkable virtue that they could have been produced in the early universe as thermal relics. Once in kinetic and chemical equilibrium, neutralinos have weak-scale pair annihilation cross sections which are such that, after they decouple from the thermal bath, their thermal relic abundance could be in the right ball park for them to be the main cold dark matter constituent. The dark matter would then be yet another relic from the early universe, very similar to light elements or the cosmic microwave background.

The weak interaction cross sections with which neutralinos interact among themselves and with the rest of the particle content of the Standard Model are such that neutralinos can be detected: a vast array of experiments are looking for tiny energy depositions that would result from the scattering of neutralinos off nuclei, or for indirect signals from occasional pair-annihilations of neutralinos in the galactic halo.

Assuming the lightest neutralino is indeed the (stable) LSP, a crucial question -- both for collider phenomenology and for the search for particle dark matter -- is what the mass of the neutralino is or might be. If the soft supersymmetry breaking SU(2) and U(1) gaugino masses satisfy the grand-unified mass relation
\begin{equation}
M_1\ =\ \frac{5}{3}\ \tan^2\theta_W\ M_2,
\end{equation}
then the limit on the lightest chargino mass from LEP indirectly sets a lower limit on the mass of the lightest neutralino (which we will indicate hereafter with the symbol $\chi$) of around $m_\chi\gtrsim 46$ GeV. Dropping this assumption, however, and considering $M_1$ and $M_2$ as unrelated parameters, particle physics does not impose any general lower bound on the neutralino mass \cite{nomass1,nomass2}.

Cosmology, however, does set a lower bound on the mass of a stable neutralino LSP. The extension to neutralinos of the argument used in the late seventies by Lee and Weinberg \cite{lw} and, independently, by others \cite{otherlw}, to constrain the mass range of heavy neutrinos, implies that $m_\chi\gtrsim{\cal O}(10)$ GeV \cite{hooperplehn,bottino,belanger} -- the precise number being irrelevant for the present discussion. As explained in detail below, this type of argument depends on assumptions on the cosmological model: in particular, the crucial hypothesis is that the universe was radiation dominated at the time when neutralinos decoupled from the primordial thermal bath, and that entropy was conserved from that point on.

The scope of the present analysis is to relax the latter assumptions of radiation domination at neutralino freeze-out and of entropy conservation, and to study the detectability of very light neutralinos, which in a generic setup can be almost arbitrarily light and still perfectly viable dark matter candidates. While models with GeV or lighter neutralinos would be indistinguishable among themselves with the LHC (the precision in the determination of missing energy and other kinematic quantities at a hadron collider is worse than a fraction of a GeV), dark matter detection experiments, in principle, give us a handle on the determination of the particle dark matter mass. Hence it is crucial to understand whether the lightest lightest neutralinos can be detected as dark matter particles or not.

We will argue in this paper that theoretical prejudices on the mass of weakly interacting massive particles (WIMPs) fostered ultra-sensitive dark matter detection experiments in certain mass ranges which are, however, totally lacking in sensitivity to lighter particle dark matter species. As Shakespeare reminds us in the words of Hamlet (Act 1, Scene V), ``{\em There are more things in heaven and earth, Horatio, Than are dreamt of in your philosophy}'': in the opinion of who writes it is the responsibility of theorists to go beyond the principle of Occam's razor and to present all possibilities that might eventually lead to the much longed for discovery of the nature of particle dark matter. While we discuss here a specific extension to the Standard Model, the MSSM, and a specific particle dark matter candidate, the lightest neutralino, we raise a more generic question: are we ready to detect a dark matter particle with a mass outside the weak-scale range? There are several reasons to believe that, from the point of view of dark matter models, there is nothing special about the weak scale, neither phenomenologically nor theoretically (see {\em e.g.} the recent interesting discussion by Feng and Kumar, \cite{Feng:2008ya}).

In the present analysis we will argue that sub-GeV neutralinos are cosmologically and astrophysically viable dark matter particles (sec.~\ref{sec:lln} and \ref{sec:cosmo}), that GLAST could in principle detect them (sec.~\ref{sec:id}) and, finally, that in the quest for light neutralinos, experiments searching for spin-dependent dark-matter-nucleon scattering appear to be more promising than those searching for a coherent scalar interaction on heavy target nuclei (sec.~\ref{sec:dd}).

\section{The Lightest Lightest Neutralinos}\label{sec:lln}
The mass of the lightest neutralino in the MSSM can, in principle, be arbitrarily small. Phenomenologically, in the MSSM a ``light'' neutralino (say, with a mass much smaller than the mass of electro-weak gauge bosons, $m_\chi\ll m_{Z,W}$) is necessarily bino-like: higgsino- or wino-like neutralinos, due to the structure of the mass matrix of neutralinos and charginos, carry an almost degenerate (up to small corrections) chargino; a light chargino is ruled out by direct searches for chargino pair-production at LEP \cite{lep2chargino}. Another option which we won't consider here, and which doesn't change our analysis or conclusions, is a light singlino-like neutralino, in extensions of the MSSM (see {\em e.g.} Barger {\em et al.}, \cite{Barger:2005hb}, McElrath \cite{McElrath:2005bp} and Gunion {\em et al.}, \cite{Gunion:2005rw} for studies of several aspects of light singlino-like neutralino phenomenology).

Mathematically, there are two possibilities for the mass of the bino-like neutralino in the MSSM to be small: (1) a fine-tuned cancellation can lead to a vanishing eigenvalue in the neutralino mass matrix; (2) a split-supersymmetric setup is realized where $M_1\ll M_{\rm susy}$.

The first possibility, which has been considered in the past by several authors, including the recent analyses of Ref.~\cite{Gogoladze:2002xp,Barger:2005hb,Dreiner:2007fw,Langenfeld:2007pf}, leads to consider the condition of a vanishing determinant for the neutralino mass matrix:
\begin{eqnarray}\label{eq:detzero}
{\rm Det}({\mathcal M}_{\chi^0})&=&\mu\Big(M_2\sin\left(2\beta\right)\sin^2\theta_W+  \\
\nonumber &&M_1\sin\left(2\beta\right)\cos^2\theta_W-M_1M_2\mu\Big)=0.
\end{eqnarray}
Eq.~(\ref{eq:detzero}), as noted a long time ago {\em e.g.} in Ref.~\cite{ref15gogo}, and as pointed out more recently in Ref.~\cite{Gogoladze:2002xp,Barger:2005hb}, under the assumption of unification of gaugino masses at the GUT scale, features the solution
\begin{equation}\label{eq:condition}
\mu M_2=\frac{m_Z^2}{r}\sin\left(2\beta\right)\left(r\cos^2\theta_W+\sin^2\theta_W\right).
\end{equation}
where $r=M_1/M_2\simeq0.5$. Interestingly, this points to a {\em positive} sign for the combination $\mu M_2$, as favored by the apparent deviation of the muon anomalous magnetic moment data from the SM expectation \cite{dam}, as well as by the inclusive $b\rightarrow s\gamma$ decay \cite{bsg}, as remarked in \cite{Barger:2005hb}. Unfortunately, however, Eq.~(\ref{eq:condition}) points to at least one light chargino, and is in conflict with LEP searches \cite{lep2chargino}, unless $\tan\beta$ is very close to unity \cite{Gogoladze:2002xp}, which, in turn, is in conflict with Higgs searches at LEP \cite{lep2chargino}.

Relaxing the GUT-scale gaugino mass unification assumption, Eq.~(\ref{eq:detzero}) can be satisfied for finely tuned values of the bino soft SUSY breaking mass $M_1$ if
\begin{equation}\label{eq:condit2}
M_1=\frac{M_2m_Z^2\sin\left(2\beta\right)\sin^2\theta_W}{\mu M_2-m_Z^2\sin\left(2\beta\right)\cos^2\theta_W}
\end{equation}
Since, as observed above, limits on the chargino mass indicate that $\mu M_2\gg m_Z^2$, we get the approximate relation (rigorously valid in the large $M_2$ limit)
\begin{equation}\label{eq:approxcondit}
M_1\simeq\sin\left(2\beta\right)\sin^2\theta_W\frac{m_Z^2}{\mu}\simeq1.9\ {\rm GeV} \ \frac{2\tan\beta}{1+\tan^2\beta}\left(\frac{\rm TeV}{\mu}\right).
\end{equation}

A second possibility is to again resort to non-universal gaugino masses, but, instead of tuning $M_1$ to satisfy the condition expressed in Eq.~(\ref{eq:condit2}), to consider the $M_1\rightarrow 0$ limit (or, rather, the $M_1/m_Z\ll1$ limit). Due to the structure of the neutralino mass matrix, and to the lower bound on $M_2$ from limits on the chargino mass, the wino component of the lightest neutralino is entirely negligible in the small $M_1$ limit. Reducing the eigenvalue problem to a third order algebraic equation and expanding the smallest eigenvalue in powers of the small parameter $M_1/\mu$, we get a mass for the lightest neutralino of
\begin{eqnarray}\label{eq:largemass}
\nonumber &m_\chi\simeq\sin\left(2\beta\right)\sin^2\theta_W\frac{m_Z^2}{\mu}
\simeq1.9\ {\rm GeV} \ \frac{2\tan\beta}{1+\tan^2\beta}\left(\frac{\rm TeV}{\mu}\right)&\\
&[M_1/m_Z\ll1\ \ {\rm limit}].&
\end{eqnarray}
This means that, for large $\tan\beta$, the lightest neutralino gets a mass of the order of 1 keV for $\mu\sim10^4$ TeV.

A light neutralino from a small value of $M_1$ and a large value of $\mu$ might seem finely tuned or unnatural, but this needs to be contrasted with the fulfillment of Eq~(\ref{eq:condit2}), which also requires a fine tuning of the order of $m_\chi/M_{\rm susy}$. The small $M_1$ solution is reminiscent of a split-supersymmetric scenario \cite{splitsusy}, although extremely large values for the soft supersymmetry breaking scalar masses are here not strictly necessary (beyond the usual limits imposed by supersymmetric contributions to the muon anomalous magnetic moment, flavor physics or collider searches).
\begin{figure*}
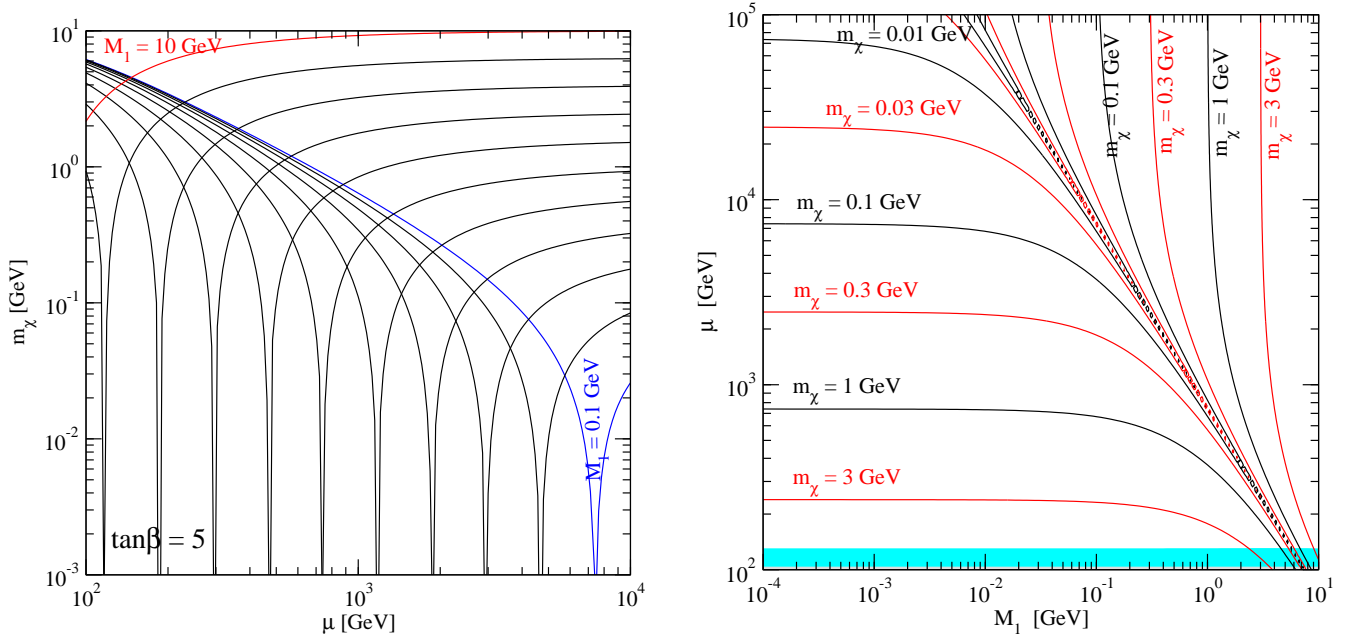

\mbox{\epsfig{file=massmu05.eps,width=8.5cm,angle=0}\qquad \epsfig{file=mass05.eps,width=8.5cm,angle=0}}
\caption{(Left): The lightest neutralino mass as a function of $\mu$, for logarithmically evenly spaced values of $M_1$ between 0.1 and 10 GeV, and for $\tan\beta=5$. (Right): Lightest neutralino mass contours in the $(M_1,\mu)$ plane, for $\tan\beta=5$; the light blue region is excluded by data on the invisible width of the $Z$ \cite{invwidth}, while the barely sivisble light gray region around $\mu\sim100$ GeV is ruled out by chargino searches at LEP \cite{lep2chargino}.}
\label{fig:mass05}
\end{figure*}
\begin{figure*}
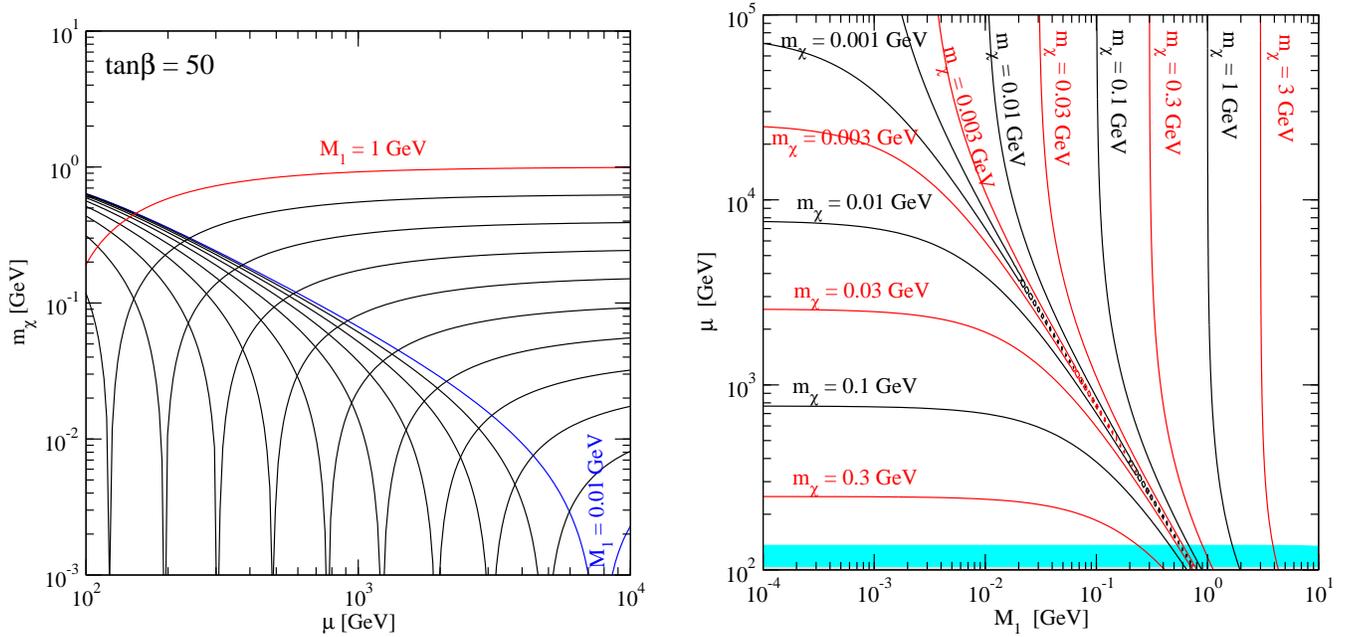

\mbox{\epsfig{file=massmu50.eps,width=8.5cm,angle=0}\qquad \epsfig{file=mass50.eps,width=8.5cm,angle=0}}
\caption{As in fig.~\ref{fig:mass05}, but for $\tan\beta=50$. In the left panel, the values of $M_1$ range here between 0.01 and 1 GeV.}
\label{fig:mass50}
\end{figure*}
Armed with the theoretical understanding outlined above, we now embark in the endeavor of scanning the relevant parameter space in search of the lightest lightest neutralinos, as a prelude to the study of their cosmology as dark matter particles (sec.~\ref{sec:cosmo}) and the possibility of their detection (sec.~\ref{sec:id}, \ref{sec:dd}).

As pointed out above, for low mass neutralinos the wino component is always negligible. Also, as far as the lightest neutralino mass and composition is concerned, soft supersymmetry breaking scalar masses are similarly unimportant, as long as radiative contributions to the neutralino mass matrix are comparatively small, as they typically are in the MSSM \cite{loopmssm}, or at least in phenomenologically viable realizations of it. In this respect, the relevant parameter space for the investigation of light neutralino masses is restricted to three parameters: $M_1$, $\mu$ and $\tan\beta$.

We will detail in the following sections \ref{sec:id} and \ref{sec:dd} how the precise pattern of the supersymmetric mass spectrum affects the phenomenology of light neutralinos as dark matter candidates. For the moment, we set all supersymmetric masses to large values. While the precise mass scale is unimportant for the computation of the lightest neutralino mass, we give here an account of the supersymmetric mass spectrum we use. We set the SU(2) and SU(3) gaugino soft breaking masses $M_2$ and $M_3$ to 1 TeV, the mass of the CP odd Higgs $M_A=1$ TeV, all the trilinear scalar couplings to zero, and all the soft supersymmetry breaking sfermion masses to 2 TeV. In addition, we entirely neglect CP violating phases. We call this approach ``{\em conservative}'', as lighter sfermions, or a lighter wino, or a lighter Higgs mass spectrum can enhance certain quantities related to the detectability of light neutralino dark matter. Below, we will also consider ``{\em optimistic}'' setups where soft supersymmetry breaking masses are assumed to be smaller, but still compatible with collider searches, precision measurements and other constraints. 

For the purpose of illustrating cases where the neutralino mass is asymptotically small, we restrict our scan to the parameters $M_1$, $\mu$ and $\tan\beta$ and refer to the above outlined conservative models for the rest of the supersymmetric particle spectrum. In particular, we set in fig.~\ref{fig:mass05} $\tan\beta=5$, and show curves at constant neutralino mass on the $(M_1,\mu)$ plane. The left panel shows the values of the lightest neutralino mass as a function of $\mu$, for a series of evenly spaced, on a logarithmic scale, values of $M_1$, ranging from $M_1=0.1$ GeV (blue line) to 10 GeV (red line). Two features discussed above emerge from the left panel. First, the dips corresponding to a zero of the neutralino mass matrix determinant are clearly visible, and their location corresponds to solutions to Eq.~(\ref{eq:approxcondit}). Secondly, the envelope of the curves for low values of $\mu$ exactly reproduces the lightest neutralino masses predicted by Eq.~(\ref{eq:largemass}). Notice that for large values of $\mu$ the lightest neutralino is purely bino-like and its mass asymptotically saturates the relation $m_\chi\simeq M_1$.

The right panel of fig.~\ref{fig:mass05} refers to the same models of the left panel, conservative plus $\tan\beta=5$, and illustrates the curves at constant neutralino mass on the $(M_1,\mu)$ plane. The most striking feature is the location of the zeros of the neutralino mass matrix determinant, which exactly matches the condition expressed by Eq.~(\ref{eq:approxcondit}). Secondly, in the low $M_1$ limit the reader can verify that the values of the neutralino mass perfectly agree with the prediction of Eq.~(\ref{eq:largemass}).

In the lower part of the right panel (small $\mu$), we shade in gray the region of parameter space where the chargino is too light, and in light blue the one where the invisible width of the $Z$ is in conflict with the current experimental limit from LEP \cite{invwidth}. Since the latter depends solely on the combination $|Z_{13}|^2-|Z_{14}|^2$ \cite{loopmssm} (see eq.~\ref{eq:zneut} below), which in turn, for light neutralinos, scales as $m_Z^2/\mu^2$, the constraint we get from the invisible $Z$ width is almost independent from the value of $M_1$ or the lightest neutralino mass.

In fig.~\ref{fig:mass50} we carry out the same exercise outlined above, this time for the case of conservative models with a large value of $\tan\beta=50$. The same comments apply, and everything scales according to the $\sin(2\beta)$ factor which appears on both Eq.~(\ref{eq:approxcondit}) and (\ref{eq:largemass}). Notice that it is generically easier (in the sense of less fine tuning being needed at a given neutralino mass) to obtain very low mass neutralinos for large $\tan\beta$.

\section{Cosmology and Astrophysics of Light Neutralinos}\label{sec:cosmo}

One of the attractive features of weakly interacting dark matter is that particle dark matter might result from the same fundamental mechanism that lies at the foundation of the success of Big Bang theory: the decoupling of species from thermal equilibrium. Primordial nucleosynthesis of light elements and the decoupling of the cosmic microwave background radiation are outstanding examples. The speculation that the same mechanism -- the freeze-out of particles from thermal equilibrium at some point in the early universe -- also lead to the production of particle dark matter is motivated by the observation that indeed particles with electro-weak scale mass and interactions might give rise to a thermal relic abundance similar to the inferred cold dark matter density. This quantitative fact is sometimes dubbed the ``{\em WIMP miracle}'', even though, as pointed out recently in Ref.~\cite{Feng:2008ya} a combination of masses and couplings giving raise to a thermal relic abundance close to the dark matter density might in principle be unrelated to the weak scale itself. In any event, this argument relies on the tacit assumption that the universe was radiation dominated at the time when the dark matter effectively decoupled from the thermal bath and that the entropy of matter and radiation is conserved after decoupling.

The nature of the universe prior to Big Bang Nucleosynthesis (BBN) is not quantitatively known. We observationally know, however, that the universe was radiation dominated at least after the onset of the synthesis of light elements, which occurred, roughly, at temperatures around $T\sim 1$ MeV. 

It is nonetheless meaningful to ask the question: which constraints on particle dark matter models are implied by the assumption that the universe was radiation dominated at particle dark matter decoupling? Also, it is equally interesting to study the implications on particle dark matter models of enforcing that all of the dark matter is made up by a single particle species produced via the mechanism of thermal decoupling. While it would be overly presumptuous to try to give a fair account of these types of studies, even for special particle physics models, such as supersymmetry, it is useful to summarize here a few facts and recent results. 

Quite generically \cite{kt}, the thermal relic abundance of any weakly interacting fermion, in a radiation dominated universe, is larger than the critical energy density of the universe unless the fermion mass is smaller than about 0.1 keV or is larger than about 2-5 GeV (the precise value of these numbers depends upon various particle physics inputs, such as the Dirac or Majorana nature of the fermion, or the number of effective relativistic degrees of freedom at the particle's freeze-out). The first constraint is usually referred to as the Cowsik-McClelland limit \cite{cmc}, while the second as the Lee-Weinberg limit \cite{lw, otherlw}. In essence, the thermal relic abundance of light (say $m_\chi\lesssim 1$ MeV) fermions grows proportionally with mass, very insensitively to the details of the fermion interaction with other particles in the thermal bath; the particle decouples when it is still relativistic, and the resulting dark matter scenario is referred to as {\em hot dark matter}. On the other hand, more massive particles decouple when they are only mildly realistic, or non-relativistic ({\em warm} or {\em cold} dark matter). In this case the computation of the particle relic abundance is more involved, but generically, for particles with a given effective Fermi-type pair annihilation cross section the thermal relic abundance scales with the inverse of the particle mass squared. At a given mass, the thermal relic abundance of weakly interacting fermions is bounded from below. This bound increases with decreasing masses and, as stated above over-closure occurs, roughly, if the particle is lighter than around 2-5 GeV.

The considerations above apply to the case of the lightest neutralino as a dark matter candidate, as recognized long ago {\em e.g.} in the seminal studies of Cabibbo {\em et al.}, \cite{earlydm1}, Pagels and Primack \cite{earlydm2} and Goldberg \cite{goldberg}. The lightest neutralino can be either super-light, a fraction of a tenth of a keV, or relatively massive, more massive than a proton.

An important piece of further observational information came from the realization that the formation of structures in a hot dark matter universe, due to the free-streaming of relativistic particles, proceeds in a top-down fashion: superclusters of galaxies form first, and later fragment into galaxies and clusters. This type of cosmology is in conflict with several observations, including the degree of homogeneity in the matter distribution at large scales and the power spectrum of the cosmic microwave background radiation. The currently accepted picture is, instead, that if dark matter is a thermally produced relic its mass cannot be below a keV or so. Ref.~\cite{abazi}, \cite{selj} and \cite{viel1,viel2} recently put this statement on firm and quantitative grounds, showing that the damping of small scale density fluctuations below the free-streaming scale of light, thermally produced dark matter particles is strongly constrained by observations of small scale cosmological structure. The most stringent bounds arise from observations of the clustering of gas along the line of sight to distant quasars: the density fluctuations of the gas follow that of the dark matter to the scale where the gas becomes pressure supported. The most recent of the mentioned studies, using a combination of cosmic microwave background observations, Sloan Digital Sky Survey Lyman-$\alpha$ forest data and the 3D power spectrum of galaxies, and a high-resolution HIRES data set puts a 95 C.L. limit on the mass of a generic thermal relic at 4 keV \cite{viel2}.

In short, the lightest neutralino in the MSSM can therefore (1) be the main dark matter constituent and (2) be thermally produced in a radiation dominated universe if and only if its mass exceeds a few GeV. It is interesting to ask what precisely is the lowest mass the lightest neutralino can have in the MSSM under the two assumptions above. This question was addressed in a few recent papers, including Hooper and Plehn \cite{hooperplehn}, Bottino {\em et al.} \cite{bottino} and Belanger {\em et al.} \cite{belanger}. These studies point to a lowest neutralino mass around 6 GeV. Although we will not consider here extensions to the MSSM, we mention that when for example gauge singlet superfields are added to the theory, loopholes can appear in the above considerations, including resonant annihilations of the dark matter particle with a light singlet Higgs. We refer the reader to recent related studies \cite{Barger:2005hb,McElrath:2005bp,Gunion:2005rw} for extensive discussions of light neutralinos in non-minimal extensions of the Standard Model.

As outlined in the Introduction, in the present study we assume that the lightest MSSM neutralino is the dark matter (in the sense that it is its dominant contributor, in terms of energy density), but we drop the assumption that the universe was radiation dominated prior to primordial nucleosynthesis.

The dilution of unwanted relics from the early universe has a long and illustrious story \cite{kt}. The gravitino and the monopole problems were some of the earliest motivations for inflation \cite{earlyinfl}. As recognized in several studies, late episodes of entropy production (or, equivalently, scenarios with low-temperature reheating) can do the same job that the old inflationary models were envisioned to be doing for relics produced near the grand unification or the Planck scale: they can dilute -- and also, possibly, non-thermally produce -- relics with an otherwise excessive relic abundance. An inexhaustive list of examples of such cosmological scenarios includes models with moduli decay \cite{moduli}, Q-ball decay \cite{qball} and low-reheating thermal inflation \cite{theinflation}. The common denominator to all these models is a ``late'' episode of entropy production when the universe is reheated to a temperature $T_{\rm RH}$ at which, effectively, radiation domination starts. The reheating temperature may have been as low as 4 MeV \cite{rhlimit1} taking into account all available cosmological data, while information from Big Bang Nucleosynthesis alone places a lower bound around 2 MeV \cite{rhlimit2} if active neutrino oscillations are taken into account, and up to around 0.7 MeV \cite{rhlimit3} if oscillations are neglected.

Giudice {\em et al.} remarked in Ref.~\cite{giudice} that low reheating scenarios can lead to keV Standard Model neutrinos being viable candidate warm dark matter particles, provided the reheating temperature is around an MeV. A systematic and prototypical study of cosmological models where the early universe is dominated by the energy density of fields  that subsequently decay and give rise to the radiation dominated era, and of the implications of such models for the relic abundance of WIMPs, was carried out in Ref.~\cite{gond1,gond2}. In their study, Gelmini {\em et al.} consider the simplest case of a single scalar field $\phi$ featuring a mass $m_\phi$ a decay width $\Gamma_\phi$ and producing on average $b$ WIMPs per decay. They point out that the two relevant parameters in this cosmological setting can be considered to be the following: the ratio $b/m_\phi$ and the reheating temperature $T_{\rm RH}$, which in turn is a function of $\Gamma_\phi$. The main conclusion of \cite{gond1,gond2} is that even in the context of such a simplified setup, the relic abundance of almost any MSSM neutralino can be brought into accord with the inferred dark matter density. For instance, Ref.~\cite{gond2} explicitly shows how with a $T_{\rm RH}\sim100$ MeV models featuring a ``standard'' relic abundance $\Omega_\chi h^2\gtrsim 10^4$ can actually produce, with a small enough value for $b/m_\phi$, relic abundances smaller than $10^{-4}$ in the modified cosmological setup. An even lower reheating temperature produces a larger suppression of the relic abundance \cite{gond2}.

A necessary condition for the dilution of the relic density is, however, that the freeze-out temperature $T_{\rm f.o.}$ is higher than the reheating temperature. As mentioned above, the latter needs to be -- conservatively -- larger than $\sim4$ MeV \cite{rhlimit1}. A crucial question we therefore need to address in studying the cosmology of the lightest lightest neutralinos is whether the freeze-out temperature is or not above 4 MeV. If it is, the results of \cite{gond1,gond2} assure us that there are viable cosmological setups that turn models with almost any thermal relic abundance into viable dark matter scenarios. In what follows, we first outline the models for light neutralinos we will employ in the remainder of this analysis; we will then discuss the range for the ``standard'' relic abundance and for the freeze-out temperatures for those models, and argue that in essentially all cases $T_{\rm f.o.}>T_{\rm RH}$, implying that these models can be brought into accord with the inferred cold dark matter density with a suitable low temperature reheating episode. We then close this section commenting on astrophysical constraints on very light neutralinos resulting from Supernov\ae.

As described in the previous section, we consider here two conservative models, where we keep as free parameters $M_1$ and $\mu$, fix $\tan\beta=5$ and 50, and set all other soft supersymmetry breaking parameters to a large (1 or 2 TeV) scale, so that all collider and other phenomenological constraints are automatically satisfied \cite{Langenfeld:2007pf}, as long as the lightest neutralino coupling to the $Z$ is sufficiently suppressed. This, in turn, forces a lower limit on $\mu\gtrsim 150$ GeV (see the right panels of fig.~\ref{fig:mass50} and \ref{fig:mass05}). 

We contrast this setup to one where, instead, we set all soft supersymmetry breaking parameters to the smallest phenomenologically viable values. Specifically, we set $M_2=150$ GeV (to be conservatively consistent with LEP2 searches for charginos), $M_A=200$ GeV (compatibly with constraints on the $b\to s\gamma$ branching ratio), soft breaking squark masses to 500 GeV and soft breaking slepton masses to 150 GeV (this values make the sfermion masses consistent with collider searches). An exception is the selectron mass, which, to avoid bounds from supernova data for very light neutralino (see below) is set to 1.2 TeV. The gluino mass is entirely irrelevant for the phenomenology under investigation here, and is set to 1 TeV. CP violating phases and trilinear couplings are assumed to vanish. As for the conservative models, we keep $\tan\beta$, $M_1$ and $\mu$ as free parameters. 

We do not however set the above spectrum in stone: for every set of values for our free parameters, we check for all available phenomenological constraints and when needed we modify the supersymmetric spectrum accordingly. Constraints from processes like those discussed {\em e.g.} by McElrath in \cite{macelrath} or those summarized by Langenfeld in \cite{Langenfeld:2007pf} can be relaxed by lifting the masses associated to certain soft supersymmetry breaking terms, with a minimal impact on the phenomenology to be discussed in this paper (the lightest neutralino relic abundance, and its direct and indirect detection). For example, excessive contributions to the muon anomalous magnetic moment $(g-2)_\mu$ in the large $\tan\beta$ regime are suppressed by invoking heavier second generation slepton masses. This does not affect significantly either the resulting light neutralino dark matter relic abundance or the dark matter direct or indirect detection rates. 

We indicate the setup outlined above, with the lightest phenomenologically viable supersymmetric spectrum, as an ``{\em Optimistic limit}''. This setup gives an indication of, for instance, upper limits to detection rates, or lower limits on the relic abundance and the freeze-out temperature. Contrasting this optimistic scenario with the conservative setup provides a guideline to understanding how important the entire supersymmetric spectrum is for the phenomenology of the lightest lightest neutralinos, compared to the choice of the values of $M_1$ and $\mu$.

\begin{figure*}
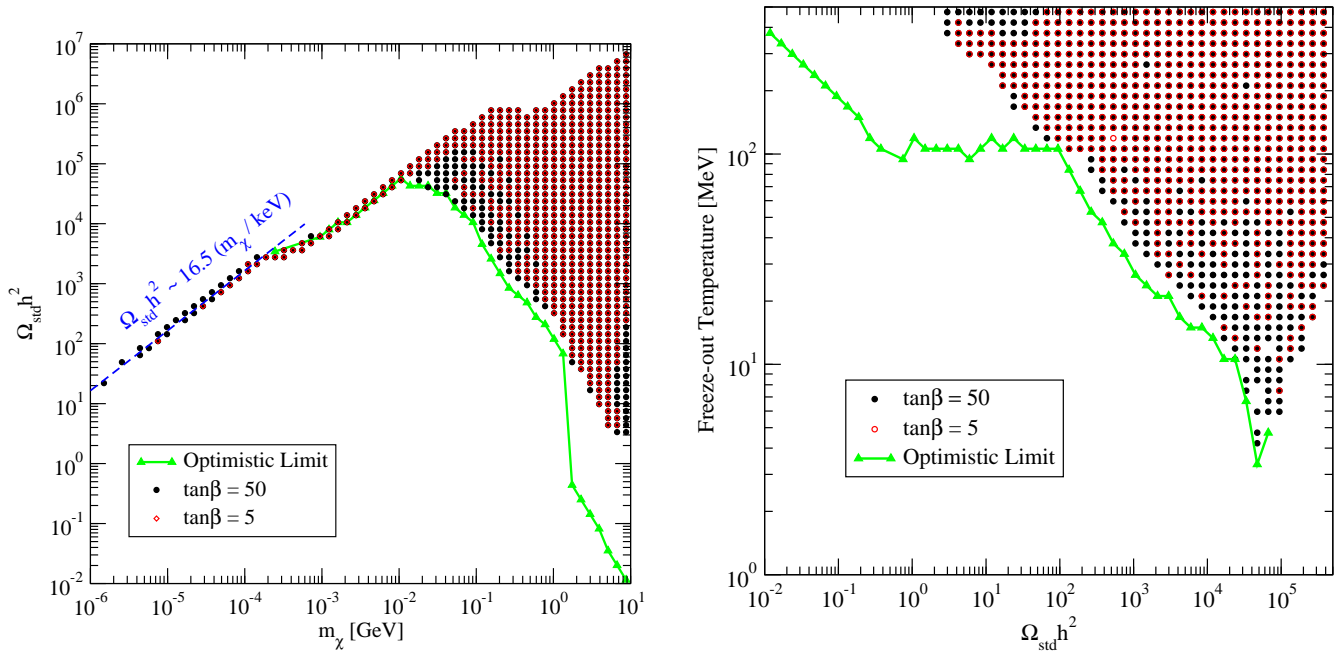

\mbox{\epsfig{file=oh2_mod.eps,width=8.5cm,angle=0}\qquad \epsfig{file=oh2tfo_mod.eps,width=8.5cm,angle=0}}
\caption{(Left): the thermal relic abundance of the lightest neutralino, as a function of its mass, in a standard cosmological scenario where the energy density of the universe is radiation dominated at arbitrarily large temperatures. In the scatter plot, filled black dots refer to the conservative models with $\tan\beta=50$, while empty red diamonds to $\tan\beta=5$. The green curve indicates ``{\em optimistic}'' models with a light supersymmetric spectrum. (Right): The correlation between the thermal relic abundance and the freeze-out temperature (defined in Eq.~(\ref{eq:tfo}), in a standard cosmological scenario) for the same models as in the left panel.}
\label{fig:oh2}
\end{figure*}

The left panel of fig.~\ref{fig:oh2} illustrates the values of the lightest neutralino mass and its relic abundance for models obtained in a scan of the $(M_1,\mu)$ plane for the two conservative cases ($\tan\beta=5,\ 50$) and for the optimistic case (with $\tan\beta=50$). We compute the relic neutralino abundance with the DarkSUSY package \cite{ds}. The result we obtain was expected on generic theoretical grounds (see {\em e.g.} fig.~5.2 in \cite{kt}): below a few MeV the freeze-out occurs when the lightest neutralino is still relativistic, and the neutralino number density per comoving volume is constant and equal to its equilibrium value. In this case, the relic number density is greatly insensitive to the details of freeze-out, and the present day relic mass density is simply given by \cite{kt}
\begin{equation}
\Omega_\chi h^2\simeq7.83\times 10^{-2}\left(\frac{g_{\rm eff}}{g_{*S}(T_{\rm f.o.})}\right)\left(\frac{m_\chi}{\rm eV}\right),
\end{equation}
where $g_{\rm eff}$ is equal to the number of internal degrees of freedom $g$ for a bosonic species, and to $3g/4$ for a fermionic species, $g_{*S}$ stands for the effective number of entropic degrees of freedom, evaluated here at the freeze-out temperature. We indeed confirm the formula above, and find, in the present case, for small masses, that 
\begin{equation}
\Omega_\chi h^2\simeq 16.5\left(\frac{m_\chi}{\rm keV}\right).
\end{equation}
The particles cease to be hot (or warm) relics ({\em i.e.} their freeze-out no longer occurs when the particles are relativistic, or mildly relativistic) for $m_\chi\gtrsim 10$ MeV. Above that mass scale, the details of particle freeze-out become important, and therefore we find an increasingly large scatter of possible relic densities. Generically, the relic abundance is bounded from below by the fact that the particle is weakly interacting, and therefore one expects a scaling of the pair annihilation cross section as $\sigma\propto E^2=m_\chi^2$. For cold relics, the thermal relic abundance is inversely proportional to the pair annihilation cross section, hence the lower bound we find approximately falls as the inverse WIMP mass squared. Eventually, for the optimistic case, it reaches the level of the cold dark matter abundance, for masses in the few GeV range, as expected from Lee-Weinberg type arguments. This is also the result of the dedicated studies mentioned above that assessed the smallest possible neutralino mass that could have the ``right'' thermal relic abundance \cite{hooperplehn,bottino,belanger}.

Since large scale structure data constrain the mass of a fermionic dark matter species to values larger than a few keV, it is clear (as also remarked long ago \cite{earlydm2}) that only few GeV neutralinos are dark matter particles compatible with a standard cosmological scenario.

The next question is whether low mass neutralinos can be accommodated as dark matter candidates relaxing the assumption of radiation domination prior to BBN. This means computing the freeze-out temperature of low mass neutralino models. One way of quantitatively defining the freeze-out temperature was proposed by Gondolo and Gelmini in Ref.~\cite{Gondolo:1990dk}, and is the one we employ in the present analysis and in fig.~\ref{fig:oh2}. Indicating with $Y(T)$ the comoving particle abundance and with $Y_{\rm eq}(T)$ the equilibrium comoving density, the freeze-out temperature is defined as the particular value of the temperature such that
\begin{equation}\label{eq:tfo}
\frac{Y(T_{\rm fo})-Y_{\rm eq}(T_{\rm fo})}{Y_{\rm eq}(T_{\rm fo})}=\delta,\qquad \delta=1.5.
\end{equation}
In our scan, we find, as expected, that the class of models giving us the ``Optimistic limit'' provide the lowest freeze-out temperatures: a lighter supersymmetric spectrum naturally induces a more efficient coupling of the lightest neutralino to the thermal bath ({\em i.e.} a larger pair-annihilation cross section), and therefore chemical equilibrium is maintained at lower and lower temperatures. For the conservative models, the freeze-out temperature essentially depends upon the coupling of the neutralino to the $Z$. In this regime, where the pair annihilation is dominated by $Z$ boson exchange into light fermions, it is easy to obtain an estimate of the decoupling temperature, as also pointed out by Ellis {\em et al.} in Ref.~\cite{Ellis:1983ew}:
\begin{equation}\label{eq:foapprox}
T_{\rm fo}\simeq\frac{1.8\ {\rm MeV}}{\left(|Z_{14}|^2-|Z_{13}|^2\right)^{2/3}},
\end{equation}
where the $Z_{1i}$ notation refers to the composition of the lightest neutralino in terms of gauge eigenstates (here, we use the conventions and notation of Ref.~\cite{Edsjo:1997hp}): 
\begin{equation}\label{eq:zneut}
\chi_1=Z_{11}\widetilde{B}^0+ Z_{12}\widetilde{W}^0+Z_{13}\widetilde{H}^0_1+Z_{14}\widetilde{H}^0_2.
\end{equation}
Since, in the limit of small $M_1$, 
\begin{equation}
|Z_{14}|^2-|Z_{13}|^2\simeq\frac{m_Z^2}{\mu^2}\ \sin^2\theta_W\ \cos2\beta,
\end{equation}
Eq.~(\ref{eq:foapprox}) implies that in the pure $Z$ exchange pair-annihilation regime one has
\begin{equation}\label{eq:tfoest}
T_{\rm fo}\simeq5.5\ {\rm MeV}\ \left(\frac{\mu}{100\ {\rm GeV}}\right)^{4/3}.
\end{equation}
In turn this means that the criterion of having a freeze-out temperature larger than the limit imposed by BBN and other cosmological observations (around 4 MeV), is automatically satisfied by a lightest neutralino interacting with ordinary matter through $Z$ exchange only: in fact, since the higgsino-like chargino is very close in mass to $\mu$, and it needs to have a mass larger than the LEP limit (around 100 GeV), then Eq.~(\ref{eq:tfoest}) dictates that $T_{\rm fo}$ is always larger than at least 5 MeV. This constitutes a proof of principle that the cosmological relic abundance of arbitrarily light neutralinos can be brought into accord with the cold dark matter density, provided the supersymmetric spectrum is such that the dominant annihilation mode is through $Z$ exchange. We remind the reader that this latter condition means, essentially, that scalar fermions are relatively heavy (for light neutralinos, the Higgs exchange modes are suppressed by the Yukawa couplings to light fermions in the final state).

The results of our scan confirm the rough estimate of Eq.~(\ref{eq:tfoest}), and indicate that in almost all light neutralino models the freeze-out temperature is in excess of a few MeV. Therefore, according to the quoted results of \cite{gond1,gond2} simple modifications to the pre-BBN cosmology can bring the relic abundance of light lightest neutralinos into accord with the observed cold dark matter abundance for arbitrarily light neutralino masses.

As pointed out above, the mass of light neutralinos is however constrained by Lyman-$\alpha$ forest data to exceed a few keV; in addition, light neutralinos can seed energy loss mechanisms in stars and Supernov\ae, especially if they are weakly but sufficiently coupled to ordinary matter. Especially stringent constraints come from data on the neutrino flux from SN1987A: neutralino emission can in principle be an effective energy loss mechanism and alter significantly the predictions for the flux of supernova neutrinos. Limits on light neutralinos and the supersymmetric particle spectrum from SN1987A neutrino data were originally proposed by Sciama \cite{sciamapreprint}. The same limits were then first studied by Ellis {\em et al.} \cite{Ellis:1988aa}, subsequently re-evaluated by Grifols {\em et al.} \cite{Grifols:1988fw,Grifols:1989qm}, by Lau \cite{Lau:1993vf}, and, more recently, re-considered by Dreiner {\em et al.} in Ref.~\cite{Dreiner:2003wh}. We will here summarize these results, and argue that they can be, in general, evaded when weak priors are put on the supersymmetric particle spectrum. 

If neutralinos have a mass comparable to the core temperature of a supernova, $T_c\sim{\cal O}(30)$ MeV, they can be produced in large numbers by $e^+e^-$ annihilation and nucleon-nucleon ``neutralino-strahlung'':
\begin{eqnarray}
e^+e^-&\to&\chi\chi\\
NN&\to& NN\chi\chi.
\end{eqnarray}
Neutralinos are then thermalized and emitted as black-body radiation from a ``neutralino-sphere'' much like the neutrino-sphere in the ``standard'' supernova model. The depth and temperature of the neutralino-sphere increase when the neutralino-nucleon elastic scattering cross section is decreased, by increasing the scalar fermion masses, resulting in an increase in the neutralino energy emission rate, which, eventually, may become comparable to the maximal rate allowed by neutrino flux measurements by the Kamiokande and IMB collaborations. For sufficiently large scalar fermion masses, instead, neutralinos can free-stream freely out of the core and the neutralino emission rate may again become acceptably small. Ellis {\em et al.} \cite{Ellis:1988aa} found that for very light neutralinos these two regimes implied that squark masses between 60 GeV and 2.5 TeV are excluded, the lower and upper bound arising in the two regimes of trapping and free-streaming, respectively. The more recent study of Dreiner {\em et al.} \cite{Dreiner:2003wh}, instead, finds that neutralino masses below 100 MeV are excluded for selectron masses between 300 and 900 GeV, while no lower bound on the neutralino mass is found for selectron masses above 1.2 TeV. The limits on squark masses quoted in Ref.~\cite{Dreiner:2003wh} are circumscribed to a very narrow range between 300 and 400 GeV. As long as the supersymmetric model is such that the scalar fermion masses are outside the mentioned ranges, then the limits from supernova 1987A do not apply. 

\section{Indirect Detection}\label{sec:id}

Among indirect detection techniques, the search for the loop-suppressed two-photon annihilation mode $\chi\chi\to\gamma\gamma$ is particularly valuable: it is hard to imagine an astrophysical process that could yield a monochromatic line that could be confused with a gamma-ray line produced by WIMP pair annihilations. This is in contrast with other indirect dark matter search channels, including the search for an excess antimatter or continuum gamma-ray emission, where the identification of a dark matter signature would be hard to disentangle from the relatively poorly understood astrophysical background.

The drawback of searching for the monochromatic pair annihilation photon line is the faintness of the expected signal: the pair annihilation cross section into two photons is suppressed by a factor of order $\alpha^2$ compared to the total pair annihilation cross section. A scan over the models under consideration here indicates that the branching ratio into $\gamma\gamma$, for neutralino masses above the pion mass threshold, ranges typically between $10^{-5}$ and $10^{-3}$. For smaller masses, the branching ratio can be larger, since the annihilation in fermionic final states is p-wave suppressed for Majorana fermions, as first emphasized by Goldberg \cite{Goldberg:1983nd}. Below the $e^+e^-$ threshold, the branching ratio is effectively 1, since the prompt annihilation mode into neutrinos is almost entirely negligible.

\begin{figure*}
\mbox{\epsfig{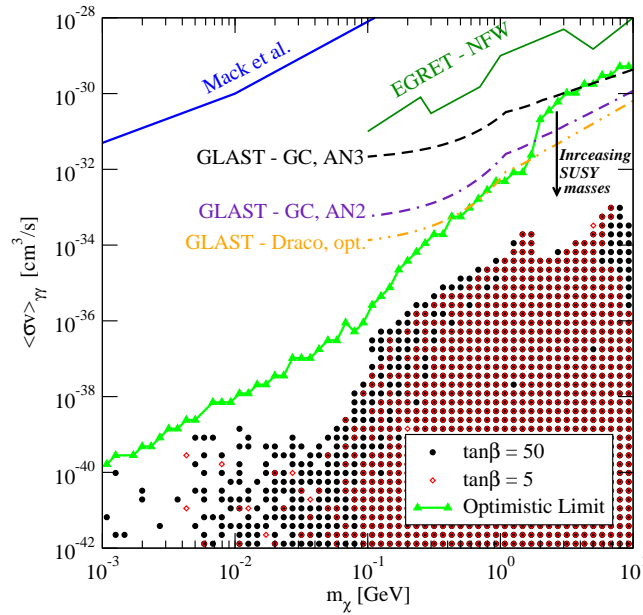}}
\caption{The correlation between the neutralino mass and its pair annihilation cross section into two photons. We also show with a solid blue line the limits derived by Mack {\em et al.} in Ref.~\cite{Mack:2008wu} and the limits derived by Pullen {\em et al.} in Ref.~\cite{Pullen:2006sy} from the EGRET data from the galactic center for a Navarro, Frenk and White profile. Dashed and dot-dashed lines indicate the anticipated performance of GLAST in searches for a line from the direction of the galactic center and from that of the Draco dwarf spheroidal galaxy.}
\label{fig:gg}
\end{figure*}
In fig.~\ref{fig:gg} we compare the $\tan\beta=5,\ 50$ conservative cases to their optimistic counterparts where the SUSY spectrum is taken to be as light as direct collider searches currently allow, on the plane defined by the neutralino mass and the pair annihilation cross section into two photons. On the same plane, we report a few observational limits. In particular, the solid blue line in the upper left corner of the figure roughly locates the limit quoted by Mack {\em et al.}, \cite{Mack:2008wu}. This refers to both INTEGRAL-SPI and COMPTEL-CGRO searches for lines from the galactic center region, and resorts to a conservative approach - for instance using the total observed gamma-ray flux as an upper limit to the dark matter signal, and making use of bin sizes typically larger than the experimental energy resolution. Less conservative assumptions were made by Pullen {\em et al.} in \cite{Pullen:2006sy} to derive a limit on the pair annihilation line from the EGRET gamma-ray data from the galactic center. The line we show refers to a NFW dark matter profile \cite{NFW}. 

We also include estimates of the sensitivity of GLAST to the monochromatic gamma-ray signal, in the direction of the galactic center and of the dwarf spheroidal galaxy Draco. For the galactic center, following \cite{Zaharijas:2006qb}, we consider two extreme choices for the background extrapolation at lower energies, as inferred from the HESS data \cite{Aharonian:2004wa} below the instrumental energy threshold. Namely, we use models 2 and 3 of Aharonian and Neronov, Ref.~\cite{Aharonian:2004jr}, (we refer to the two models as A-N2 and A-N3), respectively giving the smallest and the largest extrapolated background levels. Model A-N2 invokes inelastic proton-proton collisions of multi-TeV protons in the central super-massive black-hole accretion disk, while model A-N3 results from curvature and inverse Compton radiation. 

We assume a power-law form for the gamma-ray background,
\begin{equation}
\frac{{\rm d}\phi_b}{{\rm d}E_\gamma}=\phi_0\left(\frac{E_\gamma}{1\ {\rm GeV}}\right)^{-\gamma},
\end{equation}
and set $\phi_0\simeq1\times10^{-9}\ {\rm cm}^2{\rm s}^{-1}{\rm GeV}^{-1}$ and $\gamma=2.0$ for model A-N2, while $\phi_0\simeq3\times10^{-7}\ {\rm cm}^2{\rm s}^{-1}{\rm GeV}^{-1}$ and $\gamma=2.75$ for model A-N3. We consider the sensitivity of GLAST after five years of data taking time, assuming an average angular sensitivity of $\Delta\Omega\simeq9\times10^{-5}$ sr, and an average effective area $A_{\rm eff}$ of 5000 cm${}^2$ \cite{glastsens}. We consider a putative energy bin centered around the location of the gamma-ray line, $E_\gamma=m_\chi$, and as wide as the expected energy resolution of GLAST, $\Delta E/E\simeq 0.1$. As for the dark matter density distribution, we consider the halo profile proposed by Moore {\em et al.} in \cite{moore}. We define a signal as ``{\em detectable}'' provided the number of signal events in the considered energy bin $N_s$ is larger than 5, {\em and} if the signal is detected with a significance over background in excess of 5-$\sigma$.

We also consider the case of Draco, a nearby dwarf spheroidal galaxy, an ideal, background free and dark matter dominated environment \cite{draco}. The estimated background is only given in this case by the galactic and extra-galactic diffuse gamma-ray background, which we take to be $\phi_0\simeq6.3\times10^{-11}\ {\rm cm}^2{\rm s}^{-1}{\rm GeV}^{-1}$ and $\gamma=2.1$. We follow the results of Ref.~\cite{Colafrancesco:2006he} as far as the estimate of the dark matter integrated line of sight density squared. Taking into account the possibility that Draco hosts a central black-hole, and accounting for the adiabatic accretion of dark matter in a central ``{\em spike}'' \cite{Colafrancesco:2006he} we obtain the sensitivity reach, for GLAST, shown by the double-dotted dashed orange line. This indicates that Draco might actually be an even better target than the center of our own galaxy in the search for this kind of signal. We refer the reader to Ref.~\cite{Ferrer:2006hy} for more details.

The normalization of the GLAST sensitivity lines has a significant uncertainty stemming from the poor knowledge of the details of the innermost part of the galactic dark matter profile of the Milky Way and of Draco. The assumptions we made for Draco are certainly a best case scenario, and the absence of a spike would shift the orange line in fig.\ref{fig:gg} up possibly by orders of magnitude. The point we wish to make here is that the detection of a monochromatic gamma-ray line from very light neutralinos in satellite dwarf galaxies is, at least in principle, possible. While optimistic on the dark matter profile, our estimates are conservative as far as the understanding of the gamma-ray background is concerned. The accurate determination of the galactic and extra-galactic gamma-ray background, and the detailed understanding of the TeV gamma-ray source in the Galactic Center can make the constraints from GLAST on the gamma-ray line much tighter than what shown. Again, we do not claim that light neutralinos {\em will be} detected by GLAST in the monochromatic gamma-ray channel: simply, this is an open possibility.

\begin{figure*}
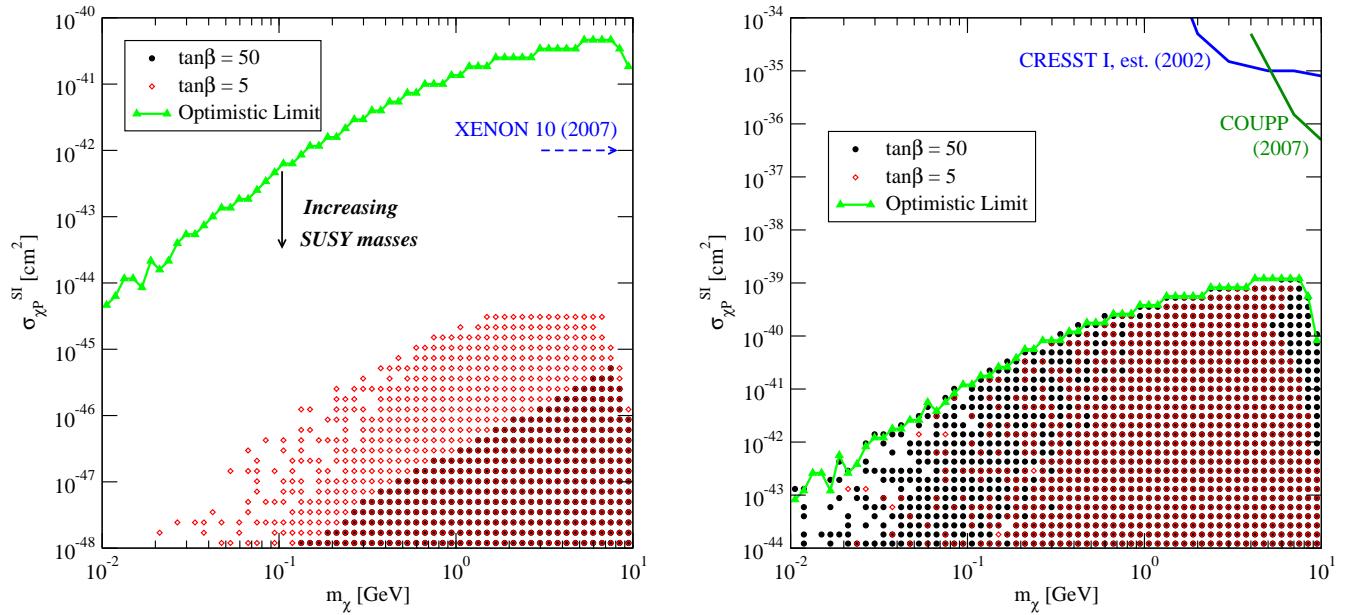

\mbox{\epsfig{file=dirdetsi_mod.eps,width=8.5cm,angle=0}\qquad \epsfig{file=dirdetsd_mod.eps,width=8.5cm,angle=0}}
\caption{The spin-independent (left) and the spin-dependent (right) scattering cross section of neutralinos off protons, as a function of the neutralino mass. We also indicate the sensitivity achieved by the XENON 10 experiment at $m_\chi=10$ GeV (left), and the limits from CRESST I and COUPP (right).}
\label{fig:dd}
\end{figure*}
Our results indicate that GLAST will not be able to detect a monochromatic gamma-ray signal from light neutralinos for our conservative benchmark models. However, as the supersymmetric spectrum gets lighter and lighter, and scalar fermions as well as the heavy Higgs sector start to contribute to the one-loop pair annihilation cross section into two photons, a few models might indeed be detectable both from the galactic center and from Draco. We find that the models which could potentially give a detectable signal have masses larger than 1 GeV. We thus conclude that the detection of sub-GeV neutralinos via the search for a monochromatic two-photon gamma-ray signal does not seem plausible with GLAST; light neutralinos with masses between 1 and 10 GeV could instead give a detectable signal.

\section{Direct Detection}\label{sec:dd}

The direct detection of dark matter is based on the idea of looking for collisions of dark matter particles with nuclei in low background environments. The field of direct dark matter detection has recently succeeded in achieving major progress in sensitivity and in demonstrating the feasibility of scaling up the size and performance of current experiments. In addition, several different techniques are being successfully explored, including solid state, noble gas and bubble chamber detectors. 

The most general possible interaction between a nucleus and a WIMP includes, in the non-relativistic limit, a coherent spin-independent coupling (scaling as $A^2$, where $A$ is the target nucleus atomic number) plus a spin-dependent coupling. In general, theory indicates that spin-independent scattering on large $A$ nuclei is the most promising direct dark matter detection technique for weak scale mass WIMPs. The CDMS-II and Xenon-10 collaborations -- employing two different target nuclei -- recently reported comparable and extremely competitive new limits on the WIMP-nucleon spin-independent scattering cross section \cite{cdmsII,xenon}. To take advantage of the coherent spin-independent scattering effect both experiments employ massive target nuclei. While this allows indeed a better sensitivity for massive WIMPs (the peak sensitivity is around a WIMP mass of 30 GeV for XENON10 and of 60 GeV for CDMS II, and scaling linearly in the WIMP mass for more massive candidates, as a result of the scaling of the local dark matter number density $\sim\rho_{\rm DM}/m_{\rm DM}$), this constitutes a strong limitation in the search for very light neutralinos. In fact, the recoil rate per unit detector mass and unit time is proportional to an integral over the WIMP velocity distribution truncated at a threshold velocity which, in turn, depends on the detector energy threshold $E_{\rm th}$ as
\begin{equation}
v_{\rm th}=\sqrt{\frac{(m_\chi+M)^2\ E_{\rm th}}{2m_\chi^2 M}},
\end{equation}
where $M$ is the target nucleus mass. In the regime where $m_\chi\ll M$ the threshold velocity $v_{\rm th}\propto \sqrt{M E_{\rm th}}$: larger nuclei have a larger velocity threshold and, with fixed WIMP velocity distribution, are effectively only sensitive to massive WIMPs. In practice, XENON10 and CDMS are barely sensitive to WIMPs below 10 GeV or so.

In fig.~\ref{fig:dd} we indicate the spin independent (left) and spin-dependent (right) WIMP-proton scattering cross section for light neutralinos. As before, we indicate the optimistic upper limit on this scattering cross section with a green curve. First, we notice that the spin-dependent cross section is very insensitive to the details of the supersymmetric spectrum, as the processes responsible for the coupling of WIMPs with nuclear spin are dominantly mediated by $Z$ boson exchange: squark exchange is only subdominant, so that the optimistic limit essentially coincides with the upper limit one gets when squarks, heavy Higgses and other scalar fermions are super-heavy. Squark contribution can instead be significant for spin independent processes, and enhance the cross section, with respect to the case when the supersymmetric spectrum is heavy with the exception of the $\mu$ parameter, by several orders of magnitude.

In the left panel we indicate the level of the XENON10 sensitivity for a WIMP mass of 10 GeV, around $10^{-42}\ {\rm cm}^2$: clearly, for light enough squarks, we obtain cross sections which are well above that level. However, as mentioned above, for standard assumptions on the WIMP velocity distribution the sensitivity of experiments employing heavy target nuclei is very poor for light dark matter candidates. In practice, XENON10 only marginally constrains the light WIMPs we are discussing here. Ideally, future direct detection experiments should be built keeping in mind that the possibility of light WIMPs is theoretically well motivated and deserves experimental attention.


Target materials in spin-dependent experiments are generically lighter than in spin-indepedent experiments, since no coherent enhancement is expected for spin-dependent recoil rates from larger nuclei. The situation looks therefore, in principle, more promising for light neutralino detection: the CRESST-I experiment \cite{cresst}, that used sapphire (${\rm Al}_2{\rm O}_3$), containing the light $O$ nucleus, has set limits on WIMPs with a very low nuclear recoil threshold (around 0.6 keV, to be contrasted with {\em e.g.} 10 keV for CDMS-II), but with a small exposure of only 1.5 kg-day (CDMS-II in contrast had an effective exposure of 121.3 kg-day). The CRESST-I sensitivity is however around four orders of magnitude away from the cross sections we predict here. The Chicagoland observatory for underground particle physics (COUPP), a bubble chamber-type experiment, which employs the target superheated liquid CF${}_3$I, reported a competitive limit on spin-dependent couplings \cite{coupp}, which we also show in fig.~\ref{fig:dd}. These limits are still between 2 and 3 orders of magnitude above the projected cross sections. However, interestingly, the current 2-kg chamber is being upgraded to 20 and 60-kg chambers, reported to be in advanced stages of construction: a scale up in the experimental sensitivity with this technique is therefore expected in a short time-scale. Spin-dependent searches therefore appear to be the most promising direct detection search channel for light neutralinos.
\vspace*{0.7cm}

\section{Summary and Discussion}
We remarked that within the MSSM neutralinos can be viable dark matter candidates even if their mass is well below a GeV, as long as certain assumptions on cosmology are relaxed: the thermal decoupling occurs typically well above the lowest possible reheating temperature allowed by BBN. Low reheating temperatures can thus dilute the thermal dark matter density of models with a large standard relic abundance down to the observed cold dark matter density in much the same way inflation is thought to have diluted excess gravitinos or topological defects in the very early universe. In this paper we asked the question whether the lightest lightest neutralinos (or, in general, weakly interacting particles at or below the GeV scale) can be detected with current dark matter search experiments. We showed that GLAST might be able, for some supersymmetric models, to detect the monochromatic gamma-ray line resulting from the pair annihilation of light neutralinos into two photons. We pointed out that despite the very significant recent progress in the field, spin-independent direct dark matter detection, which employs massive target nuclei, is not suitable to searches for light dark matter candidates. Spin-dependent direct detection appears instead to be more promising, and future experiments might be able to start to probe very light dark matter particles.

While the focus of the present analysis was on dark matter detection, with the LHC starting operations later this year we wish to briefly comment on the sensitivity that can be expected from future colliders for a scenario with very light neutralinos. The measurement of supersymmetric particle masses at the LHC will not be a straightforward task. The presence of several simultaneous supersymmetric particle production mechanisms and cascade decay chains, as well as hadronic debris from the initial state and the lack of invariant mass reconstruction due to missing transverse energy greatly complicate precision mass measurements at the LHC \cite{colliders}. Detailed information on mass differences can be gained from specific event topologies, giving raise to end-points in invariant mass distributions. The simultaneous correlation of several different mass edges can give information on the supersymmetric particle mass reconstruction \cite{lhc1,Weiglein:2004hn}. Special decay chains can also provide, under fortuitous circumstances on the supersymmetric particle spectrum, direct information on the particle masses \cite{lhc1,Bachacou:1999zb}. In addition, the overall supersymmetric mass scale of squarks and gluinos can be reconstructed from the distribution of the effective mass $M_{\rm eff}$ introduced in \cite{meff}, or from the introduction of suitably defined variables, such as e.g. the so-called $m_{T_2}$ \cite{mt2}.

In any case, the determination of the lightest neutralino mass at the LHC will greatly depend upon the details of the supersymmetric mass spectrum. To understand whether the LHC can provide information on the light neutralinos discussed here one needs to refer to a particular setup. For instance, considering the rather optimistic SPS1a benchmark \cite{Allanach:2002nj,Weiglein:2004hn}, where several supersymmetric particles will be abundantly produced at the LHC, one can expect a mass determination for the lightest neutralino with an accuracy of up to 5 GeV for a $\sim100$ GeV LSP. This stems from measurements of the kinematic endpoints of lepton spectra produced in the cascade decays of the next-to-lightest neutralino to sleptons to the LSP \cite{meff}. Mass differences are expected to be measured with an accuracy of a few GeV. Therefore, it is reasonable to expect that some information on neutralinos with a mass of $\sim10$ GeV can be gained at the LHC with an optimistic supersymmetric mass spectrum. At most, however, we will acquire information on the order of magnitude of the lightest lightest neutralino mass. In turn, this can be useful information for direct detection experiments, which would be triggered to focus on a smaller mass range, as well as for indirect detection.

With an $e^+e^-$ linear collider, instead, precision particle mass reconstruction can be carried out at an impressive level. Provided charged particles are within kinematic reach of the collider, the combination of threshold scans and the study of kinematic end-points \cite{Allanach:2002nj,Weiglein:2004hn} will provide us with accuracies down to fractions of GeV on the reconstruction of particle masses. Several studies indicated that the relative error on SUSY masses can be pushed (for instance again for the SPS1a benchmark) down to less than the 0.1\% level \cite{Weiglein:2004hn}. One can reasonably expect to have a rather accurate mass determination even for very light neutralinos, provided that the LSP mass is above say 0.1 GeV. Interestingly, this means that the limitations in the mass reconstruction at a linear collider coincide with the range of neutralino masses that might be explored with dark matter detection experiments. In turn, this implies that if a dark matter signal is detected and associated to very light neutralinos, a linear collider will be able to provide complementary information on the particle mass and on its properties. In short, our results indicate that neutralinos lighter than the mass reconstruction potential of a linear collider will not be detectable with dark matter direct or indirect detection experiments.

\section*{Acknowledgments}
The author wishes to thank Tesla Jeltema, Howard Haber, Yasunori Nomura and Piero Ullio for valuable discussions.

\end{document}